\documentclass[nohyper,12pt,letterpaper]{JHEP3}
\usepackage[dvips]{epsfig}
\usepackage{amsfonts,amssymb}

\def\ben{\begin{equation}}
\def\een{\end{equation}}

\let\a=\alpha    

\let\C=\Chi

\def\nn{\nonumber} \def\bd{\begin{document}} \def\ed{\end{document}}
\def\ds{\documentstyle} \let\fr=\frac \let\bl=\bigl \let\br=\bigr
\let\Br=\Bigr \let\Bl=\Bigl
\let\bm=\bibitem
\let\na=\nabla
\let\pa=\partial \let\ov=\overline
\newcommand{\be}{\begin{equation}}
\newcommand{\ee}{\end{equation}}
\def\ba{\begin{array}}
\def\ea{\end{array}}
\def\ft#1#2{{\textstyle{{\scriptstyle #1}\over {\scriptstyle #2}}}}
\def\fft#1#2{{#1 \over #2}}
\def\del{\partial}
\def\vp{\varphi}
\def\sst#1{{\scriptscriptstyle #1}}
\def\oneone{\rlap 1\mkern4mu{\rm l}}
\def\td{\tilde}
\def\wtd{\widetilde}
\def\ie{{\it i.e.\ }}
\def\eg{{\it e.g.\ }}
\def\dalemb#1#2{{\vbox{\hrule height .#2pt
        \hbox{\vrule width.#2pt height#1pt \kern#1pt
                \vrule width.#2pt}
        \hrule height.#2pt}}}
\def\square{\mathord{\dalemb{6.8}{7}\hbox{\hskip1pt}}}
\newcommand{\ho}[1]{$\, ^{#1}$}
\newcommand{\hoch}[1]{$\, ^{#1}$}
\newcommand{\bea}{\begin{eqnarray}}
\newcommand{\eea}{\end{eqnarray}}
\newcommand{\ra}{\rightarrow}
\newcommand{\lra}{\longrightarrow}
\newcommand{\Lra}{\Leftrightarrow}
\newcommand{\tr}{{\rm tr} }
\newcommand{\Tr}{{\rm Tr} }
\def\0{{\sst{(0)}}}
\def\1{{\sst{(1)}}}
\def\2{{\sst{(2)}}}
\def\3{{\sst{(3)}}}
\def\4{{\sst{(4)}}}
\def\5{{\sst{(5)}}}
\def\6{{\sst{(6)}}}
\def\7{{\sst{(7)}}}
\def\8{{\sst{(8)}}}
\def\n{{\sst{(n)}}}
\def\cA{{{\cal A}}}
\def\cF{{{\cal F}}}
\def\tV{\widetilde V}
\def\tW{\widetilde W}
\def\tH{\widetilde H}
\def\tE{\widetilde E}
\def\tF{\widetilde F}
\def\tA{\widetilde A}
\def\im{{{\rm i}}}
\def\tY{{{\wtd Y}}}
\def\ep{{\epsilon}}
\def\vep{{\varepsilon}}
\def\R{\rlap{\rm I}\mkern3mu{\rm R}}
\def\bD{{{\bar D}}}

\def\R{\rlap{\rm I}\mkern3mu{\rm R}}
\def\bD{{{\bar D}}}
\def\R{{{\mathbb R}}}
\def\C{{{\mathbb C}}}
\def\H{{{\mathbb H}}}
                   
\def\CP{{{\mathbb C}{\mathbb P}}}
\def\RP{{{\mathbb R}{\mathbb P}}}
\def\Z{{{\mathbb Z}}}
\def\bA{{{\mathbb A}}}
\def\bB{{{\mathbb B}}}
\def\bC{{{\mathbb C}}}
\def\bR{{{\mathbb R}}}
\def\bD{{{\mathbb D}}}
\def\bE{{{\mathbb E}}}
\def\bZ{{{\mathbb Z}}}
\def\bI{{{\mathbb I}}}
\def\Re{{{\frak{Re}}}}
\def\Im{{{\frak{Im}}}}
\def\cosec{{\,\hbox{cosec}\,}}
\def\Gm{{\Gamma_{\!\! -}}}
\def\Gp{{\Gamma_{\!\! +}}}
\def\stan{{standard }}
\def\nonstan{{supernumerary }}

\def\cosech{{\hbox{cosech}}}

\def\etcyc{{\hbox{and cyclic}}}
\def\btheta{{\bar\theta}}
\def\alp{{\alpha'}^3}

\newcommand{\tamphys}{\it Center for Theoretical Physics,
Texas A\&M University, College Station, TX 77843, USA}

\newcommand{\mitchell}{\it George P. \& Cynthia W.
Mitchell Institute for Fundamental Physics,\\
Texas A\&M University, College Station, TX 77843-4242, USA}
\newcommand{\umich}{\it Michigan Center for Theoretical Physics,
University of Michigan\\ Ann Arbor, MI 48109, USA}
\newcommand{\upenn}{\it Department of Physics and Astronomy,
University of Pennsylvania, Philadelphia,  PA 19104, USA}
\newcommand{\SISSA}{\it  SISSA-ISAS and INFN, Sezione di Trieste\\
Via Beirut 2-4, I-34013, Trieste, Italy}

\newcommand{\newton}{\it Isaac Newton Institute for Mathematical
Sciences,\\
20 Clarkson Road,  University of Cambridge,
Cambridge CB3 0EH, UK}

\newcommand{\ihp}{\it Institut Henri Poincar\'e\\
  11 rue Pierre et Marie Curie, F 75231 Paris Cedex 05}

\newcommand{\damtp}{\it DAMTP, Centre for Mathematical Sciences,
 Cambridge University\\  Wilberforce Road, Cambridge CB3 OWA, UK}
\newcommand{\itp}{\it Institute for Theoretical Physics, University of
California\\ Santa Barbara, CA 93106, USA}

\newcommand{\imperial}{\it The Blackett Laboratory,
Imperial College London\\
Prince Consort Road, London SW7 2AZ. }

\newcommand{\icrea}{\it Instituci\'o Catalana de Recerca i
Estudis Avan\c cats,\\
Departament ECM, Facultat de F\'\i sica, \\
Universitat de Barcelona, Diagonal 647,\\
E-08028 Barcelona, Spain.}

\newcommand{\auth}{
H. L\"u\hoch{\ddagger1},
C.N. Pope\hoch{\ddagger1}, K.S. Stelle\hoch{\star2} and
P.K. Townsend\hoch{\clubsuit3}}

\preprint{MIFP-03-22\\Imperial/TP/03-04/5\\UB-ECM-PF-03/32\\ 
hep-th/0312002}

\title{\Large\bf Supersymmetric Deformations of $G_2$ Manifolds from 
Higher-Order
Corrections to String and M-Theory}

\author{ H. L\"u and C.N. Pope\thanks{Research supported in part by 
DOE grant
DE-FG03-95ER40917}\\
\mitchell}

\author{K.S. Stelle\thanks{Research supported in part by the EC under 
TMR contract
HPRN-CT-2000-00131 and by PPARC under SPG grant PPA/G/S/1998/00613}\\
\imperial}

\author{P.K. Townsend\thanks{On leave from DAMTP, University of 
Cambridge, 
UK.}\\
\icrea}

\abstract{
 The equations of 10 or 11 dimensional supergravity admit 
supersymmetric
compactifications on 7-manifolds of $G_2$ holonomy, but these 
supergravity
vacua are deformed away from special holonomy by the higher-order
corrections of string or M-theory. We find simple expressions for
the first-order corrections to the Einstein and Killing spinor 
equations
in terms of the calibrating 3-form of the leading-order $G_2$-holonomy
background. We thus obtain, and solve explicitly, systems of 
first-order
equations describing the corrected metrics for most of the known  
classes
of cohomogeneity-one 7-metrics with $G_2$ structures.
}

\preprint{MIFP-03-22\\ Imperial/TP/03-04/5\\UB-ECM-PF-03/32\\ 
\tt{hep-th/0312002}}

\begin{document}

\section{Introduction}

The low-energy limits of string/M-theory are supergravity theories in
ten or eleven dimensions. These supergravity theories admit solutions
of the form (Minkowski)$_d \times K_n$, where $K_n$ is a Ricci-flat
space of dimension $n$, and $d+n$ is equal to the total spacetime
dimension. When $K_n$ is compact, a solution of this type can be
interpreted as a Kaluza-Klein vacuum, which is supersymmetric if $K_n$
admits covariantly-constant spinors.  Put another way, supersymmetric
Kaluza-Klein vacua arise if $K_n$ is a space having an appropriate
{\it special holonomy}.  The most studied case in string theory is
when $n=6$, and $K_6$ is a Ricci-flat K\"ahler, \ie Calabi-Yau (CY),
manifold. The most natural analogue in M-theory arises for $n=7$, in
which case one is interested in 7-manifolds $K_7$ with the exceptional
holonomy $G_2$. Of course, one can also consider $G_2$
compactifications in the context of string theory, and we shall take
this point of view initially, returning to M-theory in the final
Section.

 Supergravity is just the leading term in an effective action
obtained, in the case of string theory, by integrating out the
massive-modes of the string. The effective action is thus an expansion
in derivatives or, equivalently, in powers of the inverse string
tension $\alpha'$. If we focus on the part of the effective action
relevant to graviton scattering amplitudes then the $n$th term in its
expansion has the form $(\alpha')^{n-1} R^n$ where $R$ stands for the
Riemann tensor. This term has no effect on tree-level amplitudes with
less than $n$ external particles, so we would need to consider at
least the n-point amplitudes to determine it; alternatively, it can be
determined by an $n$-loop computation of the beta-function for the
associated (1,1)-supersymmetric sigma-model. At the tree-level in
string perturbation theory the massive modes of the string contribute
only to amplitudes with at least four external gravitons, so $R^2$ and
$R^3$ corrections may be absent. They {\it are} absent for the type II
superstring theories, in agreement with the three-loop finiteness of
(1,1) supersymmetric sigma models with a Ricci-flat target space so,
in these cases, the first non-trivial correction is an $\alp R^4$
term. In the string conformal frame, this term must appear with a
factor of $e^{-2\phi}$, where $\phi$ is the dilaton, so the
graviton/dilaton part of the tree-level effective Lagrangian is
\be\label{lageff}
{\cal L} = \sqrt{-g}\, e^{-2\phi}\,  \left( R + 4(\del\phi)^2 - 
c\, \alp\, Y\right) 
\ee
for a known constant $c$, proportional to $\zeta(3)$, and  a known 
scalar $Y$
that is quartic in the Riemann tensor of the 10-dimensional 
spacetime. The
corresponding equations of motion may be written
\bea\label{eqsmotion}
R_{\mu\nu} +2  \nabla_\mu\nabla_\nu\, \phi  &=& c\, \alp\, X_{\mu\nu} 
\nn\\
\square\, \phi - 2\left(\partial\phi\right)^2 &=& {1\over2}c\alp\,  
\left(Y- g^{\mu\nu}X_{\mu\nu}\right) 
\eea
where $D_\mu$ is the standard covariant derivative, and
\be\label{xvar}
\sqrt{-g}\, X_{\mu\nu} =  e^{2\phi}{\delta \over \delta g^{\mu\nu} 
}\int\!
d^{10}x\sqrt{-g}\, e^{-2\phi}\, Y. 
\ee

    The explicit form of $Y$ was first found from a four-loop
sigma-model computation \cite{grisaru1,grisaru2}, and this result was
partially confirmed by a computation of the 4-point graviton
scattering amplitude in string theory \cite{grosswitten,grosssloan}.
There is a considerable measure of ambiguity in the choice of
expression for $Y$, since one is free to perform field redefinitions
at order $\alp$ that modify $Y$ by terms that vanish upon use of the
leading-order equations of motion.

   For our purposes, it is not necessary to know the fully 
10-dimensional
form of $Y$. As we are principally concerned with the effect of $Y$ on
supergravity solutions of the form (Minkowski)$_3\times K_7$, it would
be sufficient to restrict the Riemann tensor to be non-zero on
$K_7$. However, it is convenient to consider $G_2$ compactifications
as special cases of solutions of the form (Minkowski)$_2\times M_8$,
where $M_8$ is an 8-manifold that happens to be a product of the form
$\bR\times K_7$, and to allow the Riemann tensor to be non-zero on
$M_8$. This has the advantage that it includes the possibility that
$M_8=K_8$, where $K_8$ is an 8-manifold of Spin(7) holonomy, although
we will have little to say about this case in the present paper.  
With this
understood, let $R_{ijk\ell}$ be the components of the Riemann tensor
on $M_8$ and define $t_8$ to be the tensor such that
\be
t_8^{i_1j_1 i_2 j_2 i_3 j_3} A_{i_1j_1}A_{i_2j_2}A_{i_3j_3}A_{i_4j_4}
 = 24\left[\tr A^4 - \ft14 \left( \tr A^2\right)^2\right]
\ee
for any antisymmetric $8\times 8$ matrix $A$ with entries $A_{ij}$. 
Then 
field variables can be chosen so that $Y$ is given by
\be\label{Y} 
Y\propto  \left[ t_8^{i_1\dots i_8} t_8^{j_1\dots j_8} - \ft14
\varepsilon^{i_1\dots i_8} \varepsilon^{j_1\dots j_8} \right] R_{i_1 
i_2 j_1
j_2} \ R_{i_3 i_4 j_3 j_4} \ R_{i_5 i_6 j_5 j_6} \ R_{i_7 i_8 j_7 j_8}
\,.
\ee
As we shall see later, the choice of variables implied by (\ref{Y}) 
has
the virtue that underlying structures associated with special holonomy
backgrounds are preserved, even though the special holonomy itself is 
modified or lost under the effect of the $\alp$ corrections.  

   It was pointed out in \cite{grosswitten} that (\ref{Y}) can be
written as a Berezin integral over the 16 components of a
Grassmann-odd Majorana SO(8) spinor $\psi$. Let $\tilde\Gamma^i$
($i=1,\dots,8$) be SO(8) Dirac matrices, and $R_{ijk\ell}$ the
components of the Riemann tensor of the metric on $M_8$.  Then
\be
Y \propto \int d^{16}\psi \, \exp\left[\ft12 \left(\bar\psi\, 
\tilde\Gamma^{ij}\,
\psi\right)\left(\bar\psi\, \tilde\Gamma^{k\ell}\, \psi\right) 
R_{ijk\ell}\right].
\label{fermionform}
\ee
where $\bar\psi = \psi^\dagger$. In a basis for which the Dirac
matrices $\tilde\Gamma^i$ are real, the spinor $\psi$ is real and
$\bar\psi=\psi^T$.

   Before proceeding with our analysis of the corrections to 
special-holonomy
backgrounds, it is of interest to see what can be established in 
general.
Let us assume that the dilaton is constant to leading order, so that
\be
\phi= \phi_0 + \alp\, \phi_1. 
\ee
The  functional variation (\ref{xvar}) then yields
\be\label{varyY}
X_{ij} = \tilde X_{ij} +  \nabla^k\nabla^\ell X_{ikj\ell}
\ee
where $X_{ikj\ell}$ is a tensor cubic in the curvature, antisymmetric
within first and second index pairs and symmetric under pair 
interchange;
$\tilde X_{ij}$ is a symmetric tensor quartic in curvatures which is 
traceless in eight dimensions (as one can see from the vanishing Weyl 
weight of $\sqrt{-g}Y$). The tensor $\tilde X_{ij}$ arises from the 
variations of ``explicit'' metrics in $\sqrt{-g}Y$ (\ie metrics in 
$\sqrt{-g}$ or those used to contract curvature indices, but not 
those appearing 
inside connections). If we assume that that $\phi_1$ depends only on 
the coordinates of the
compact manifold $K_n$ of a lowest-order (Minkowski)$_d \times K_n$
supergravity solution, as is required to preserve the $d$-dimensional
Poincar\'e invariance, then the dilaton equation reduces to
\be
\label{dilaton} \square \phi_1 + {c\over2} \nabla^k\nabla^\ell
\left(g^{ij}X_{ikj\ell}\right) = {c\over2} Y.  
\ee 
The left hand
side vanishes on integration over $K_n$, so we deduce that
\be
\int_{K_n} Y=0\,.\label{ycon}
\ee
This is the same as the result that was obtained in \cite{grosswitten}
by making the assumption that $M_8$ is Ricci flat.  As we see here, 
the
assumption of Ricci flatness is not needed in order to prove 
(\ref{ycon}). 

   The condition (\ref{ycon}) is trivially satisfied if $Y=0$, which 
is
the case if $K$ allows a covariantly constant SO(8) spinor $\chi$
(with respect to the Levi-Civita connection). This is because the
integrability condition for such a spinor is
\be
R_{ijk\ell}\, \tilde\Gamma^{k\ell}\, \chi=0, 
\ee
which implies that the matrices $R_{ijk\ell} \tilde\Gamma^{k\ell}$
have a common zero eigenvalue, and hence that at least one linear
combination of the components of $\psi$ in (\ref{fermionform}) does
not appear in the integrand; the Berezin integral is then zero.
However, as emphasised in \cite{freemanpope}, the vanishing of $Y$
does not imply that there will be no correction to the Einstein
equation because these corrections depend also on the the tensor
$X_{ij}$, which vanishes only if $K$ admits at least three Killing
spinors of the same SO(8) chirality. This condition is satisfied by
all hyper-K\"ahler 8-metrics (in agreement with the ultra-violet
finiteness of supersymmetric hyper-K\"ahler sigma models) but not
otherwise. Thus, for most cases of interest there will still be
corrections due to the non-vanishing of $X_{ij}$.  In the CY case, it
was shown in \cite{freemanpope} that the metric ceases to be
Ricci-flat once these corrections are taken into account, although it
remains K\"ahler. This result is actually a special case of a more
general result; it can be shown that if $M_8$ has special holonomy, 
then
\be \tilde X_{ij}=0,
\qquad g^{ij}X_{ikjl} = g_{kl} Z \label{xtilde}
\ee 
where $Z$ is the scalar cubic in
the Riemann tensor that becomes the Euler density in dimension six. In
this case (\ref{dilaton}) is solved by 
\be\label{correctdilaton}
\phi_1 = -{c\over2}Z.  
\ee 
When this is substituted into the
Einstein equation, one finds that 
\be\label{correctedE} R_{ij} = c\alp
\left[ \nabla_i\nabla_j Z + \nabla^k\nabla^\ell X_{ikj\ell}\right].
\ee 
To establish this result in full generality requires consideration
of 8-manifolds of Spin(7) holonomy, but this involves 
additional complications that we choose to postpone to a future
publication.  In the next section we shall establish the result for 
8-manifolds of
the form $M_8= \bR\times K_7$ with $K_7$ a Ricci-flat 7-manifold of
$G_2$ holonomy; in this case we shall find that
\be\label{formofZ} X_{ikj\ell} =
{1\over2} \left[c_{ik m}\, c_{j\ell n}\ Z^{mn} + (i\leftrightarrow
j)\right] \qquad (i,j,k,\ell=1,\dots,7) 
\ee 
where $c_{ikm}$ is the associative 3-form on the $G_2$ manifold, 
and   
$Z^{mn}$ is a
symmetric tensor on $K_7$ that arises in the analysis of this case. We
shall show that this incorporates the Calabi-Yau results as a special 
case.

 From (\ref{correctedE}) we see that the string corrections must
deform the special holonomy metric in CY, $G_2$ and Spin(7)
compactifications to one that is no longer Ricci flat.  An important
question then arises as to whether the modified solution is still
supersymmetric.  To answer this question, one needs to know the
modifications to the supersymmetry transformation rules of the
fermions, in particular the gravitino, to order ${\a'}^3$. A program
to determine these modifications has been underway for some years,
\eg \cite{pvhw}, but the results obtained to date are not sufficient
for our purposes. However, it was shown in \cite{cfpss,frposost} for
the special case of CY compactifications that there exist candidate
corrections to the supersymmetry transformation of the gravitino, such
that the supersymmetry of the supergravity solution is preserved in
the face of the ${\a'}^3$ corrections. The modified Killing spinor
equation that one finds in this way naturally involves structures that
are special to CY manifolds, but it was also shown in
\cite{cfpss,frposost} that the corrected Killing spinor equation can
be written in a form that makes sense for any Riemannian manifold. We
shall show here that this modification of the Killing spinor equation
also ensures preservation of supersymmetry for modified $G_2$
compactifications to order $\alp$.

A spacetime of the form (Minkowski)$_d \times K_n$ is a solution of
the $(d+n)$-dimensional supergravity theory if the $K_n$ metric is
Ricci-flat, irrespective of whether it is compact or not. The 
compactness of
$K_n$ is needed only for the interpretation of such solutions as
Kaluza-Klein vacua; even in this case we may wish to consider a
non-compact Ricci-flat manifold that approximates a given compact
$K_n$ near a singularity, and this has the advantage that an explicit
metric may then be available. If so, one can explicitly determine the
corrections to the $K_n$ metric needed to maintain supersymmetry in
the face of the $\alp$ corrections.  This issue was addressed for some
six-dimensional cohomogeneity-one CY manifolds in \cite{froltsey}, and
more recently for these and further eight-dimensional CY manifolds in
\cite{lupost}. The same issue for $G_2$ manifolds of cohomogeneity one
was addressed in \cite{froltsey} and we take up this issue again here.

Of course, Ricci-flat manifolds of $G_2$ holonomy are of most interest
in the context of compactifications of M-theory to four-dimensional
supergravity theories with N=1 supersymmetry \cite{paptown}, and it is
the M-theory effective action that is relevant in that case. It is
known that there is a similar $R^4$ correction to the action of
11-dimensional supergravity, although it is more closely related to
the 1-loop contribution to the IIA effective action than to the
tree-level contribution, which actually has a different form. However,
this difference is irrelevant for $G_2$ compactifications, as are the
CS terms, so that our results can also be carried over to M-theory, 
as we
explain in the final Section.

\section{Corrections to $G_2$ Metrics}\label{g2sec}

Our main concern, at least initially, is tree-level string corrections
to supergravity solutions of the form (Minkowski)$_2\times M_8$, for
an 8-manifold $M_8$ of special holonomy, and a constant dilaton. Under
these circumstances, all corrections to the supergravity equations are
determined by the $M_8$ tensor
\be
\label{xvar8}
X_{ij} = {\delta \over \delta g^{ij} }\int\! d^8x\sqrt{g}\,  Y. 
\ee
where $Y$ is given by (\ref{fermionform}). From (\ref{varyY}) and 
(\ref{xtilde}) we
expect to find that
\be
X_{ij} = \nabla^k\nabla^\ell X_{ikj\ell}
\ee
and our first goal will be to show that if $M_8=\bR\times K_7$, for a
7-manifold $K_7$ of $G_2$ holonomy, then $\tilde X_{ij}=0$ and 
the tensor $X_{ijkl}$ takes the form (\ref{formofZ}).

To this end, we first write the real SO(8) Dirac matrices as
$\tilde\Gamma^i = \tilde \Gamma^{\underline i}e_{\underline i}{}^i$, 
where $\tilde\Gamma^{\underline i}$ are the constant Dirac matrices 
in a frame defined by the achtbein 
$e_{\underline i}{}^i$. As there is a naturally defined `8' 
direction, by hypothesis, we may suppose that the achtbein is block 
diagonal with its $1\times1$ block equal to unity. We may now set
\be\label{gammarep}
\tilde\Gamma^{\underline i} 
= \sigma_2 \otimes \Gamma^{\underline {i}}  \qquad {\underline 
i}=1,\dots,7; \qquad 
\tilde\Gamma^8 = -\sigma_1 \otimes \oneone_8,
\ee
where $\oneone_8$ is the $8\times 8$ identity matrix, and then define the 
antisymmetric, and {\it imaginary} $8\times 8$ SO(7) Dirac matrices 
$\Gamma^i$ ($i=1,\dots,7$) from $\Gamma^{\underline i}$ using the 
$7\times 7$ achtbein  block.  We choose the signs such that
\be\label{signs}
i\Gamma^{\underline 1}\cdots \Gamma^{\underline 7} = \oneone_8. 
\ee
Chiral SO(8)  spinors  are eigenspinors of
\be
\tilde\Gamma_9 \equiv \tilde\Gamma^{\underline 
1}\cdots\tilde\Gamma^{\underline 8} =  
\sigma_3\otimes \oneone_8.
\ee
As this matrix is diagonal, we have
\be
\psi = \pmatrix{\psi_+\cr \psi_-}
\ee
where $\psi_\pm$ are real 8-component SO(7) spinors such that $\psi$
is chiral if $\psi_-=0$ and anti-chiral if $\psi_+=0$.

The expression (\ref{fermionform}) may now be written as
\be
Y \propto \int d^8\psi _+ d^8\psi_- \, 
\exp\left[ \left(\bar\psi_+\, \Gamma^{ij}_+\,
\psi_+\right)\left(\bar\psi_-\,  \Gamma^{k\ell}_-\, 
\psi_-\right) R_{ijk\ell}\right]
\label{fermionform2}
\ee
where from (\ref{gammarep}) we have the $8\times8$ Gamma matrices
\be
\Gamma^{ij}_+ = \Gamma^{ij}_- = \Gamma^{ij} = \ft12\left[\Gamma^i,
\Gamma^j\right],
\qquad 
\Gamma^{i8}_+ = -\Gamma^{i8}_- = i\Gamma^i \qquad
(i,j=1,\dots, 7)\ .
\ee 
This $\pm$ symmetric form is actually the form of this integral originally given
in
\cite{grosswitten}. For the Dirac matrices as specified above, 
$\bar\psi_\pm =
\psi_\pm^T$.  Performing the Berezin integration one has
\bea
Y &\propto& \ep^{\a_1\cdots \a_8}\, \ep^{\beta_1\cdots \beta_8}\, 
\left[(\Gamma^{i_1 i_2})_{\a_1\a_2}\cdots 
(\Gamma^{i_7 i_8})_{\a_7\a_8}\right] \left[
(\Gamma^{j_1 j_2})_{\beta_1\beta_2}\cdots 
   (\Gamma^{j_7 j_8})_{\beta_7\beta_8}\right] \times \nn\\
  && \qquad \qquad  R_{i_1 i_2 j_1 j_2}\  R_{i_3i_4j_3j_4}\  
R_{i_5i_6j_5j_6} \  R_{i_7 i_8 j_7 j_8}\,, \label{epform}
\eea
where $\a_1,\beta_1, \dots$ are 8-component spinor indices of $SO(7)$.

   We now take the eight-dimensional transverse space to be of the 
form
$\R \times K_7$ where $K_7$ is a seven-manifold of $G_2$ holonomy.  On
$K_7$ the decomposition $G_2\subset SO(7)$ implies that the
8-dimensional spinor representations decompose as
\be
8_\pm \longrightarrow 7+1\,.\label{g2decomp}
\ee
The singlet corresponds to a covariantly constant real SO(7) spinor
$\eta$.     It is convenient to normalise this (commuting) spinor such
that $\bar\eta\eta=1$, where $\bar\eta= \eta^T$ for a pure-imaginary
representation of the $SO(7)$ Dirac matrices.     A useful Fierz 
identity is
\be
\Gamma_i\, \eta\, \bar\eta\, \Gamma_i + \eta\, \bar\eta = \oneone\,.
\label{d7fierz}
\ee

In computing the variation of $Y$ we can make use of the zeroth-order
conditions implied by the $G_2$ holonomy of the unperturbed
background, {\it after} having performed the variation, of course.
Let's first see why $\tilde X_{ij}=0$ in (\ref{varyY}) for such 
manifolds. The tensor $\tilde X_{ij}$ arises from variation of the
achtbeins used to define the Dirac matrices  $\tilde \Gamma^i$ from 
the 
constant  matrices $\tilde\Gamma^{\underline i}$ (because the 
variation    
of the $\sqrt{-g}$ factor yields a contribution proportional to $Y$ 
that vanishes
in the undeformed background). This leads to a term of the form
\be
\delta e_{\underline i}{}^i \left[\left(\bar\psi_+ 
\Gamma^{{\underline i}\, j}\psi_+\right) T^+_{ij}
+ \left(\bar\psi_- \Gamma^{{\underline i}\, j}\psi_- 
\right)T^-_{ij}\right]
\ee
in the integrand of (\ref{fermionform2}), where $T^\pm_{ij}$ are 
tensors containing the remaining $14$ independent spinors  with the 
property that they vanish  whenever any of these 14 spinors is a 
Killing  spinor.  We have two Killing spinors of opposite chirality, 
so in each of the above two terms, one of them must appear in the 
associated $T$-tensor, and hence the coefficient of the achtbein 
variation vanishes.    

In contrast, variation of the metric appearing in a curvature tensor 
produces a non-vanishing 
result since the variation now involves a variation of a $T$-tensor, 
and the varied $T$-tensor 
does not have the crucial property of vanishing whenever one of its 
$14$ spinors is a Killing spinor. 
Alternatively, this follows from the observation that a varied 
curvature need not  have special holonomy (\ie  need not have zero 
eigenspinors on either front or back indices). Using the Palatini 
identity for the variation of the curvature, one finds that
\bea
\delta Y &\propto&  \ep^{\a_1\cdots \a_8}\, \ep^{\beta_1\cdots 
\beta_8}\,
\left[(\Gamma^{i_1 i_2})_{\a_1\a_2}\cdots 
(\Gamma^{i_7 i_8})_{\a_7\a_8}\right]\left[
(\Gamma^{j_1 j_2})_{\beta_1\beta_2}\cdots
   (\Gamma^{j_7 j_8})_{\beta_7\beta_8}\right] \times \nn\\
 && \qquad\qquad R_{i_1 i_2 j_1 j_2} \ R_{i_3i_4j_3j_4} \ 
R_{i_5 i_6 j_5 j_6}\, 
\nabla_{i_7}\nabla_{j_7}\, \delta g_{i_8 j_8}\,.\label{epformvar}
\eea
where the three unvaried Riemann tensors are those of the unperturbed
background.

   An arbitrary Majorana spinor on $K_7$ decomposes according to
(\ref{g2decomp}), and from (\ref{d7fierz}) we see that $i\Gamma_i\eta$
provides a mapping between the reduced 7-component index of a spinor
that is orthogonal to $\eta$, and the 7-component vector index.
The SO(7) Clifford algebra is spanned by the real symmetric matrices
$(\oneone, i\Gamma_{ijk})$ and the real antisymmetric matrices
$(i\Gamma_i,\Gamma_{ij})$. It follows that the only independent
covariantly-constant tensor that we can form from the Killing spinor
spinor $\eta$ is
\be
\label{cdef}
c_{ijk} = \im\, \bar\eta\, \Gamma_{ijk}\, \eta\,.
\ee
This is the calibrating 3-form in the $G_2$ manifold $K_7$.  Its 
Hodge dual
\be
\label{cdual} 
c^{ijk\ell} \equiv \ft16 \ep^{ijk\ell mnp}\, c_{mnp} = \bar\eta\,
\Gamma_{ijk\ell}\, \eta
\ee
is also covariantly constant; we assume here that tensors have
components with respect to an orthonormal frame.  Using (\ref{cdef}),
(\ref{cdual}), and (\ref{d7fierz}), one can deduce that
\be
c_{ijm}\, c^{k\ell m} = - c_{ij}{}^{k\ell} + 2 \delta_{ij}^{k\ell} 
\,,\qquad
c_{ijmn}\, c^{k\ell mn} = -2 c_{ij}{}^{k\ell} + 8 
\delta_{ij}^{k\ell}\,,\qquad
\ee

The expression (\ref{epformvar}) for $\delta Y$ can now be simplified
using properties of $G_2$ manifolds. We can choose a spinor frame in
which the 8-component spinor indices decompose as $\a=(\bar\a, 8)$,
etc, where the covariantly-constant spinor $\eta^\a$ has a
non-vanishing component only when $\a=8$.  The mapping between a
7-vector and the reduced spinor $\eta^{\bar\a}$ is then implemented
with $(\Gamma_i\eta)_{\bar\a}$.  Using this mapping, and making use of
the $G_2$-manifold identity
\be
R_{ijk\ell}\, c^{k\ell}{}_{mn} = 2 R_{ijmn}\label{g2riem},
\ee
we see that $\delta Y$ can be written as
\be
\delta Y \propto  \ep^{m i_1\cdots i_6}\, \ep^{n j_1\cdots j_6} \, 
R_{i_1 i_2 j_1 j_2}\ R_{i_3i_4j_3j_4}\  R_{i_5 i_6 j_5 j_6}\, 
     c^{ij}{}_{m}\, c^{k\ell}{}_{n}\ \nabla_i\nabla_k \, \delta 
g_{j\ell} .
\ee
The constant of proportionality can be absorbed into the constant $c$
in (\ref{eqsmotion}), so an integration by parts in (\ref{xvar8})
yields the result
\be
X_{ij} = c_{ik m}\, c_{j\ell n}\, \nabla^k\nabla^\ell\, Z^{mn}\,,
\label{Xdef}
\ee
where 
\be
Z^{mn} \equiv  \ft1{32} \ep^{m i_1\cdots i_6}\, \ep^{n j_1\cdots j_6} 
\, 
R_{i_1 i_2 j_1 j_2}\cdots R_{i_5 i_6 j_5 j_6}\,.\label{zmndef}
\ee
Note that $Z^{ij}=Z^{ji}$, and that $\nabla_i\, Z^{ij}=0$ as a
consequence of the Bianchi identity for the Riemann tensor.  Note also
that 
\be
g^{ij}\, X_{ij} = \square Z\label{Xtrace}
\ee
where
\be
Z=g_{mn}Z^{mn}\,.\label{zdef}
\ee

   We thus obtain the modified Einstein equation 
\be\label{correctedg2} 
R_{ij} = c\alp \left[\nabla_i\nabla_j Z + c_{ikm}c_{j\ell n}
\nabla^k\nabla^\ell Z^{mn}\right]
\ee 
This takes the general form (\ref{correctedE}) with the tensor
$X_{ikj\ell}$ as given in (\ref{formofZ}).  The dilaton is given by
\be
\phi = -\ft12 c\, \alp\, Z\,.
\ee

If one takes $K_7= \bR\times K_6$ with $K_6$ a Calabi-Yau 6-manifold 
then the
only non-zero components of the tensor $Z^{ij}$ are $Z^{77}=Z$. As
\be
c_{ij7} = J_{ij} \qquad (i,j=1,\dots,6)\label{g7complexstr}
\ee
where $J_{ij}$ is the K\"ahler 2-form, we have
\be
R_{ij} = 
c\alp \left[\nabla_i\nabla_j  + J_{ik}J_{j\ell} 
\nabla^k\nabla^\ell\right] Z\ .
\label{kahlercase}
\ee
As $Z$ is the Euler density for $K_6$, we recover the Calabi-Yau 
result of
\cite{freemanpope}.

    Although it was not necessary for the discussion we gave above, it
is instructive to note that there exists a fully covariant Lagrangian
that gives a modified Einstein equation which reduces to
(\ref{correctedg2}) upon restriction to a leading-order 
(Minkowski)$_3$
times $G_2$-holonomy background.  Specifically, this is obtained by
expanding out the product of eight-dimensional epsilon tensors in
(\ref{Y}) in terms of Kronecker deltas, and then allowing the indices
to range over all ten values.  Terms of quadratic or higher order in
the Ricci tensor or Ricci scalar coming from this expansion are
unimportant when considering $\alp$ corrections to the leading-order
$G_2$ backgrounds, but the terms of linear order in Ricci tensor and
scalar, together of course with the quartic Riemann tensor terms, do
contribute to the order $\alp$ equations of motion in these
backgrounds.  The terms linear in the Ricci tensor and Ricci scalar 
could be removed by field redefinitions, but since we have implicitly 
made a 
specific choice of field variables, the coefficients of these linear 
terms are now uniquely determined.  Specifically, one finds that
$Y$ is given by
\be
Y= W_0 - W_1 -W_2\,,\label{yw}
\ee
where
\bea
W_0  &=& 12(  R_{a_1 a_2 b_1 b_2} \, R^{a_2 a_3 b_2 b_3}\,
R_{a_3 a_4 b_4}{}^{b_1} \, R^{a_4 a_1}{}_{b_3}{}^{b_4} -
R_{a_1 a_4 a_3 b_3}\, R^{a_2 b_2 a_3 b_3}\,
R^{a_1}{}_{b_1 b_2 b_4}\, R_{a_2}{}^{b_1 a_4 b_4})\,,\nn\\
W_1 &=& 6R^{ab}\,(4 R_{a}{}^{cde}\, R_b{}^f{}_d{}^g\,
R_{efcg} + 2 R_a{}^d{}_b{}^e\, R_{echg}\, R_d{}^{chg} -
R_a{}^{cde}\, R_{bcfg}\, R_{de}{}^{fg})\,,\label{w012def}\\
W_2 &=& R\, (R_{abcd}\, R^{cdef}\,
R_{ef}{}^{ab} - 2
      R_{acbd}\, R^{cedf}\, R_e{}^a{}_f{}^b)\,.\nn
\eea
Note that $W_2$ is nothing but $R$ times the covariant expression for
$Z$ introduced in Eqs\ (\ref{zmndef}) and (\ref{zdef}), namely
\be
Z =  R_{abcd}\, R^{cdef}\,
R_{ef}{}^{ab} - 2
      R_{acbd}\, R^{cedf}\, R_e{}^a{}_f{}^b\,.\label{zexp}
\ee

     Using the Lagrangian corrections defined by (\ref{yw}), with the 
specific
coefficients of the scheme-dependent (\ie redefinition-dependent) 
terms $W_1$ and $W_2$,
ensures that the resulting field equations yield (\ref{kahlercase}) 
if $K_7$ is chosen to
be of the form $\bR\times K_6$ (or $S^1\times K_6$). This is a 
natural choice of scheme for
this case, since it preserves the K\"ahler structure of the metric on 
$K_6$
(although it does, of course, lift the Ricci-flatness of the metric, 
implying that the
holonomy enlarges from $SU(3)$ to $U(3)$). This can be seen directly 
from the corrected
Ricci condition (\ref{kahlercase}) by writing it in holomorphic 
coordinates, in which case it
takes the form $R_{a\bar b}=c\alp\partial_a\partial_{\bar b}Z$. The 
same scheme choice
(\ref{yw}) also ensures that, starting from a general $G_2$ holonomy 
manifold $K_7$,
the modified field equation is expressible as (\ref{correctedg2}). 
Although in this case, 
unlike the K\"ahler case described above, there is no intermediate 
enlargement of the
holonomy group between $G_2$ and the full $SO(7)$, it nevertheless 
seems the natural
$G_2$ generalisation of the K\"ahler-preserving Ricci correction
(\ref{kahlercase}).\footnote{In particular, as we have just seen, it 
has the feature that
if one chooses the $G_2$ manifold to be of the form $\bR\times K_6$, 
it preserves the
K\"ahler structure of $K_6$.} As we shall next see, this form of the 
$G_2$ correction
allows, as in the K\"ahler case, an elegant modification of the 
covariant derivative
appearing in the Killing spinor equation, thus permitting 
supersymmetry preservation in the
corrected background.

\section{The Corrected Killing Spinor Equation}

We seek a modified covariant derivative $\hat\nabla_i$ such that the 
`modified
$G_2$' space $K_7$ admits a (Killing) spinor $\eta$ satisfying
$\hat\nabla_i\eta=0$ to order $\alp$.  We expect
\be\label{g2Ddef}
 \hat \nabla_i  = \nabla_i + c\alp Q_i 
\ee
where $Q_i$  acts by (matrix) multiplication on  $\eta$. Let us define
\be
Q_{ij}\equiv \nabla_i\, Q_j - \nabla_j\, Q_i. 
\ee
Then, to order ${\a'}^3$, the integrability condition
$[\hat\nabla_i,\hat\nabla_j]\, \eta=0$ gives
\be\label{intcon}
\ft14 R_{ijk\ell}\, \Gamma^{k\ell}\, \eta + c\alp Q_{ij}\, \eta =0\ .
\ee
Multiplying by (\ref{d7fierz}), we see that this is equivalent to
\be
R_{ijk\ell}\, c^{k\ell}{}_{m} + 4c\alp  \im\, \bar\eta\, \Gamma_m\, 
Q_{ij}\, \eta 
=0\,,\qquad 
\bar\eta\, Q_{ij}\ \eta =0\,.\label{intcon2}
\ee
Multiplying instead by $\Gamma_i$ and then using (\ref{intcon2}), we
also deduce that
\be
R_{ij} = 2c\alp\,  \bar\eta\, \Gamma_{jk}\, Q_i{}^k\, 
\eta\,.\label{intcon3}
\ee
Our criterion for determining $Q_i$ will therefore be that it should
satisfy $\bar\eta\, Q_{ij}\, \eta=0$ and that when substituted into
(\ref{intcon3}), it should yield the corrected field equation
(\ref{correctedg2}).

We find that the following $Q_i$ fulfils these criteria:
\be
Q_i = -\ft{\im}{2}\,  c_{ijk}\, \nabla^j\, Z^{k\ell}\, \Gamma_\ell\,.
\label{Qidef}
\ee
The verification of $\bar\eta\, Q_{ij}\, \eta=0$ is immediate, and 
from
(\ref{intcon3}) we find
\bea
R_{ij} &=& - c\alp \left( c_{jkn}\, c^{k\ell}{}_{m} 
\, \nabla_i \nabla_\ell\, Z^{mn} -
   c_{i\ell m}\, c_{jkn}\, \nabla^\ell\nabla^k\, Z^{mn}\right) \nn\\
&=& c\alp\, \left(\nabla_i\nabla_j\, Z +  c_{i\ell m}\, 
c_{jkn}\,\nabla^\ell
\nabla^k\, Z^{mn}\right)\,.
\eea
We have now found a modified Killing spinor equation such that a space
admitting a Killing spinor must solve the modified Einstein equation
to order $\alp$. One would expect this condition to be equivalent, to
the vanishing of the supersymmetry variation of the gravitino to order
$\alp$, but the latter makes sense for any background metric whereas
our expression for $Q_i$ involves the associative 3-form $c_{ijk}$
that exists only on manifolds of $G_2$ holonomy. However, it is
possible to rewrite $Q_i$ in purely Riemannian terms. To do this, we
first use (\ref{zdef}) and the identity
\be
\Gamma_i = \ft{\im}{6!}\, \ep_{i j_1\cdots j_6}\, \Gamma^{j_1\cdots 
j_6}
\label{d7gammacon}
\ee
to rewrite (\ref{Qidef}) as
\be
Q_i = \ft1{64} \, c_{ijk}\, \ep^{k i_1\cdots i_6}\, 
\nabla^j\, (R_{i_1 i_2 j_1 j_2}\cdots R_{i_5 i_6 j_5 j_6})\, 
\Gamma^{j_1\cdots j_6}\,.\label{Qi2}
\ee
Next, we note that
\bea
c_{ijk}\, \ep^{k i_1 \cdots i_6} &=& \im\, \bar\eta\, \Gamma_{ij}\, 
\Gamma_k\, 
\eta\, \ep^{k i_1\cdots i_6}\,,\nn\\
&=& -\bar\eta\, \Gamma_{ij}\, \Gamma^{i_1\cdots i_6}\, \eta\,,\nn\\
&=& 2\delta_{ij}^{[ i_1 i_2}\, 
\bar\eta\, \Gamma^{i_3\cdots i_6]}\, \eta\,,\nn\\
&=& 2 \delta_{ij}^{[i_1 i_2}\, c^{i_3\cdots i_6]}\,.
\eea
 From the property (\ref{g2riem}) we then find that (\ref{Qi2}) 
reduces
to
\be
Q_i = \ft3{16} \, \nabla^j\, (R_{ij m_1 m_2}\, R^{k\ell}{}_{m_3 
m_4}\, 
R_{k\ell m_5 m_6} - 4 R_{ik m_1 m_2} \, 
R_{j\ell m_3 m_4} \, R^{k\ell}{}_{m_5 m_6}) \, \Gamma^{m_1 \cdots m_6}
\label{Qi3}
\ee
which is indeed purely Riemannian.

After further manipulations, which involve distributing the derivative
in (\ref{Qi3}) and using the Bianchi identity, one can further show
that $Q_i$ can also be expressed as
\be
Q_i = -\ft34 \,  (\nabla^j\, R_{ik m_1 m_2})\, R_{j\ell m_3 m_4}\, 
R^{k\ell}{}_{m_5 m_6}\, \Gamma^{m_1 \cdots m_6}\,.
\ee
In this form, $Q_i$ can be recognised as precisely the same
modification to the Killing spinor condition that was proposed in
\cite{cfpss}.  In that case, the proposal was based on the
consideration of deformations from $SU(3)$ holonomy for
six-dimensional Calabi-Yau backgrounds, so there was no a priori
reason to expect the same expression in the $G_2$ case.

 \section{Explicit Examples}

 In this Section, we shall consider some examples of 7-dimensional 
metrics of
cohomogeneity-one, and, by making use of the modified Killing spinor 
equation 
\be\label{mod}
\left(\nabla_i - {i\over2}\alp c_{ijk}\nabla^j Z^{k\ell}\, 
\Gamma_\ell\right)\eta=0 \ .
\ee
Note that we set $c=1$ in this Section, as this can always be arranged
by a choice of units for $\alpha'$. We use this Killing spinor
condition to derive first-order equations for the metric coefficients
such that the metrics will give supersymmetric solutions of the 
modified Einstein equations
(\ref{correctedg2}).\footnote{Details of the uncorrected metrics and 
expressions for the tensors
$c_{ijk}$ can be found in Refs
\cite{brysal,gibpagpop,cglpnoslump,gukov}.}

\subsection{$S^3\times S^3$ principal orbits with $SU(2)^3$ symmetry}
\label{s3s31sec}

For our first example, we take the case of cohomogeneity-one metrics
with $S^3\times S^3$ principal orbits and $SU(2)^3$ isometry. This
class of metrics was considered in \cite{brysal,gibpagpop}, where it
was shown that there exists a complete, non-singular Ricci-flat
solution with $G_2$ holonomy.  One starts from the metric ansatz
\be
ds^2 = dt^2 + a^2\, (\sigma_i - \Sigma_i)^2 + b^2\, (\sigma_i + 
\Sigma_i)^2\,,
\ee
where $\sigma_i$ and $\Sigma_i$ are two independent sets of
left-invariant 1-forms for the group $SU(2)$, and $a$ and $b$ are
functions of $t$.  Since the undeformed $G_2$ metric, and hence also
the deformed one, admits a single Killing spinor $\eta$, it follows
that in the natural basis for the vielbein and the spin frame, this
spinor must be independent of the coordinates on the $S^3\times S^3$
orbits. We shall take
\be
e^0 = dt\,,\qquad e^i = a\, (\sigma_i-\Sigma_i)\,,\qquad 
e^{\hat i} = b\, (\sigma_i + \Sigma_i)\,,
\ee
from which it follows that
\bea
\omega_{0i} &=& -\fft{\dot a}{a}\, e^i\,,\qquad 
\omega_{0\hat i} = - \fft{b}{b}\, e^{\hat i} \, ,\qquad 
\omega_{ij} = (\fft{b}{4a^2} - \fft1{2b})\, \ep_{ijk}\, e^{\hat k}\, 
,\nn\\
\qquad  \omega_{\hat i \hat j} &=& - \fft{1}{4b}\, \ep_{ijk}\, e^k
\,,\qquad \omega_{i \hat j} = - \fft{b}{4a^2}\, \ep_{ijk}\, e^k\,.
\eea
After calculating $D_i$ in this frame one sees easily that $\eta$ will
be constant.

It now becomes a straightforward matter to read off the equations that
follow from requiring that such a constant $\eta$ satisfy the Killing
spinor equation (\ref{mod}).  As usual, we may substitute the
uncorrected first-order equations for $G_2$ holonomy when evaluating
the correction term $Q_i$ in (\ref{g2Ddef}), given in (\ref{Qidef}),
since we are working here only to order ${\a'}^3$, and there is
already an explicit factor of ${\a'}^3$ associated with this
term. Thus the term $Q_i$ can be expressed in purely algebraic
terms. We find the following first-order conditions:
\be
\fft{\dot a}{a} + \fft{b}{2a^2} +  \dot S_1 =0\,,\qquad
\fft{\dot b}{b} - \fft{b}{4a^2} + \fft{1}{4b} +  \dot S_2=0\,,
\label{2funfo}
\ee
where 
\bea
S_1 &=& -\alp \ft1{9216}\, 
a^{-12}\, (6851 a^6 - 38274 a^4\, b^2 + 69147 a^2\,
b^4 - 39312 b^6)\,,\nn\\
S_2 &=&\alp \ft5{9216}\,   a^{-12}\, (463 a^6  - 2892 a^4\,b^2  + 
4563 a^2\,  b^4  - 1890 b^6)\,.\label{s1s2def}
\eea
Note that the $\alp$ terms are total time derivatives. In their
absence, equations (\ref{2funfo}) reduce precisely to the standard
first-order conditions for $G_2$ holonomy, yielding the non-singular
solution obtained in \cite{brysal,gibpagpop}. We want to determine how
this solution is modified by the $\alp$ terms. Defining a new radial
variable $\rho$ by $dt = - b^{-1}\, d\rho$, and letting $A\equiv a^2$,
$B \equiv b^2$, the equations (\ref{2funfo}) become
\be
\fft{dA}{d\rho} + 2 A\, \fft{d S_1}{d\rho} 
= 1\,,\qquad \fft{1}{B}\, \fft{dB}{d\rho} - \fft1{2B} + 
\fft{df}{d\rho} =0\,,
\ee
where
\be 
f = \ft12 \log A + S_1 + 2 S_2\,.
\ee
After some simple manipulations we therefore arrive at the solution
\bea
a^2(\rho) &=& e^{-2 S_1(\rho)}\, \int_{\rho_1}^\rho e^{2 S_1(x)}\, 
dx\,,\nn\\
b^2(\rho) &=& \fft1{2 a(\rho)}\, e^{-\left[S_1(\rho) +2 
S_2(\rho)\right]}\,
\int_{\rho_0}^\rho a(x)\, e^{\left[S_1(x) + 2S_2(x)\right]}\, dx\,.
\label{absol}
\eea

Regular solutions arise if we take $\rho_1 < \rho_0$, so that as
$\rho$ runs from the asymptotic region $\rho=\infty$ downwards, the
metric function $b$ vanishes while $a$ is still non-zero.  Provided
that $\rho_1 <\rho_0$, the precise choice of value for $\rho_1$ is
arbitrary, since the system of equations has a shift symmetry
$\rho\longrightarrow \rho + {\rm constant}$.  A convenient choice is
to take $\rho_1$ to be the constant such that $a^2(\rho)$ in
(\ref{absol}) is given by
\be
a^2(\rho) = e^{-2 S_1(\rho)}\, \left[ \rho + \int_{\rho}^\infty 
\left(1-e^{2 S_1(x)}\right)\, dx \right] \,.\label{asimpsol}
\ee
The solution (\ref{absol}) to (\ref{2funfo}) would be exact if we
viewed $S_1$ and $S_2$ as arbitrarily-specified external ``source
functions.'' In fact, of course, $S_1$ and $S_2$ are themselves given
by (\ref{s1s2def}), and so (\ref{absol}) should be viewed as
integro-differential equations governing the functions $a$ and $b$.

 The equations (\ref{absol}) can easily be solved up to the order
$\alp$ to which we are working, since one can simply substitute the
leading-order solutions for the undeformed $G_2$ holonomy metric into
the expressions (\ref{s1s2def}) for $S_1$ and $S_2$, since they
already carry an explicit factor of $\alp$.  These leading-order
solutions can themselves be read off from (\ref{absol}) by setting
$S_1$ and $S_2$ to zero, yielding
\be
a^2=\rho\,,\qquad b^2=\ft13 \rho\, [1-(\rho_0/\rho)^{3/2}]\,.
\label{abg2sol}
\ee
These expressions give the standard complete, non-singular metric of
$G_2$ holonomy found in \cite{brysal,gibpagpop}. The explicit solution
with $\alp$ correction can then be obtained straightforwardly
using (\ref{absol}), where $S_1$ and $S_2$ take their Ricci-flat
background forms.  It is given by
\bea
a^2 &=& \rho\, \left(1 + \ft1{32} \alp\,\left[ \fft{10}{9\rho^3} + 
\fft{220 \rho_0^{3/2}}{63\rho^{9/2}} + \fft{221\rho_0^3}{48\rho^6} +
\fft{14\rho_0^{9/2}}{9 \rho^{15/2}}\right]\right)\,,\nn\\
b^2 &=& \ft13 \rho\, \left[1-(\rho_0/\rho)^{3/2}\right]\,\Bigg( 1+ 
\ft1{32}
\alp\,\Bigg[-\fft{4589}{2016(\rho\,\rho_0)^{3/2}} +
\fft{6611}{2016\rho^3} - \fft{863\rho_0^{3/2}}{672\rho^{9/2}}\nn\\
&&\qquad\qquad\qquad\qquad\qquad -\fft{451\rho_0^3}{288\rho^6} +
\fft{133\rho_0^{9/2}}{72 \rho^{15/2}}\Bigg]
\Bigg)\,.\label{s3s3sol}
\eea
This solution is finite everywhere from $\rho=\rho_0$ to 
$\rho=\infty$.

It should be noted that the solution (\ref{s3s3sol}) does not have a
smooth $\rho_0\rightarrow 0$ limit.  In fact, it is of interest to
look in closer detail at what happens if the parameter $\rho_0$ in the
original $G_2$ holonomy solution (\ref{abg2sol}) is taken to be zero,
implying $b^2 = \ft13 a^2$.  This corresponds to the situation where
the cohomogeneity-one metric is nothing but the cone over $S^3\times
S^3$, since the $G_2$ metric is then
\be
ds^2 = dt^2 + \ft1{12} t^2\, \Big(  (\sigma_i - \Sigma_i)^2 +
\ft13 (\sigma_i +  \Sigma_i)^2\Big)\,.\label{s3s3conemet}
\ee
The quantity in the large parentheses is the Einstein metric of weak
$SU(3)$ holonomy on $S^3\times S^3$.  The metric (\ref{s3s3conemet})
is, of course, singular at $t=0$, the apex of the cone.  If we now
take this leading-order solution, and follow the same procedure of
studying the higher-order corrections up to order $\alp$, the
calculations become very simple.  In fact with $b^2 = \ft13 a^2$, the
first-order equations (\ref{2funfo}) reduce, after substituting the
leading-order solution into the expressions for $S_1$ and $S_2$, to
\be
\dot a = {\sqrt{3}\over6} \Big(\fft{5 \alp}{24a^6} - 1\Big)\,.
\label{coneeom}
\ee
This can be solved by defining a new radial variable $r$ such that 
\be
dr = \left[(5/24)\alp\, a^{-6} - 1\right]\, dt\, ,
\ee
whereupon the metric becomes\footnote{Results for the modification of
$G_2$ holonomy metrics given in \cite{froltsey} suggested that the
singular cone-metric limits of the regular cohomogeneity-one $G_2$
metrics would not suffer $\alp$ corrections.  By contrast, we find
non-vanishing modifications in the cone-metric limits for all the
known $G_2$ holonomy examples, and indeed our results in general
differ from those in \cite{froltsey}.  The
apparent discrepancy between our results and those in \cite{froltsey}
may be due to a different choice of scheme.}
\be
ds^2 = \fft{dr^2}{\Big(1 - \ft{m^6}{r^6}\Big)^2} + 
\ft1{12} r^2\,  \Big(  (\sigma_i - \Sigma_i)^2 +
\ft13 (\sigma_i +  \Sigma_i)^2\Big)\,,\label{conedef}
\ee
where $m^2= 3 (45)^{1/3}\, \a'$. 

This metric is smooth as one descends from large $r$ to a minimum
value at $r=m$, which is at infinite affine distance. This is
suggestive of what one might expect from string theory given that
string theory generally permits conical singularities. Of course, the
metric should really be expanded in powers of $m^2$, and only the
leading, $m^6$, correction retained, since the higher-order terms in
this expansion are of the same order of magnitude as others we have
neglected. Actually, the expansion parameter here is $m^2/r^2$, on
dimensional grounds, so for any finite $m$ we would never be justified
in considering $r\sim m$; the geometry in a region of size $m$ near
what was the apex of the cone in the unperturbed, supergravity,
solution is inaccessible within the $\alpha'$ expansion.  However, it
is clear that string corrections do change the supergravity result in
this region, and it is plausible that this effect is captured by the
metric (\ref{conedef}).

\subsection{$S^3\times S^3$ principal orbits with 
            $SU(2)^2\times U(1)$ symmetry}

 For the next example, consider
\bea
ds^2 = dt^2 + a_i^2\,(\sigma_i-\Sigma_i)^2 + b_i^2 (\sigma_i + 
\Sigma_i)^2
\eea
where the $\sigma_i$ and $\Sigma_i$ are again the left-invariant 
1-forms of $SU(2)\times SU(2)$, and we split the triplet index as
$i=(1,\a)$.  The first-order equations can be expressed as the 
following
\bea
\fft{a_1'}{a_1} - \fft{a_1}{4a_3\, b_1} + \fft{a_3}{4a_1\, b_2} +
        \fft{b_2}{4a_1\, a_3} - \fft{a_1}{4a_2\, b_3} +
\fft{a_2}{4a_1\, b_3} + \fft{b_3}{4a_1\, a_2} + K_1 &=&0\,,\nn\\
\fft{b_1'}{b_1} + \fft{a_2}{4a_3\, b_1} + \fft{a_3}{4a_2\, b_1} -
\fft{b_1}{4a_2\, a_3} + \fft{b_1}{4b_2\, b_3} -
\fft{b_2}{4b_1\, b_3} - \fft{b_3}{4b_1\, b_2} + \wtd K_1 &=&0
\,,
\eea
together with the cyclic permutations of indices $1,2$ and $3$.  The
equations for the leading-order Ricci-flat metric were obtained in
\cite{cglpnoslump}.  We have worked out explicitly the result for the
$\alp$ contributions $K_i$ and $\wtd K_i$; however, they are too
complicated to present here since they involve thousands of terms.  In
\cite{gukov}, an analytic solution for a special case $a_3=a_2$ and
$b_3=b_2$ was obtained; in terms of the new radial variable $r$ 
defined by
\be
dt = -\sqrt{\fft{9(r-\ell)(r+\ell)}{4(r-3\ell)(r+3\ell)}}\, dr \equiv 
h(r)dr
\ee
we have
\bea
&& a_1=-\ft12 r\,,\qquad a_2=a_3 = \sqrt{\ft3{16} (r-\ell) (r+3\ell)} 
\nn\\
&& b_1 = \ell\,\sqrt{\fft{(r-3\ell)(r+3\ell)}{(r-\ell)(r+\ell)}}\,, 
\qquad
b_2=b_3= -\sqrt{\ft3{16} (r + \ell) (r-3\ell)}.
\eea
For this Ricci-flat $G_2$ background, we have
\bea
h\, K_1 &=& -\fft{327680 \alp\, \ell^2\, r (r^4 - r^2\, \ell^2 + 32 
\ell^4)}{
243(r^2-\ell^2)^7}\,,\nn\\
h\, K_2 &=& \fft{81920\alp\, \ell^2\,(r^6 - 14 r^2\,\ell +
35r^4\,\ell^2 - 64 r^3\,\ell^3 - 77 r^2\,\ell^4 -
66r\,\ell - 39\ell^6}{243(r-\ell)^8(r+\ell)^7}\,,\nn\\
h\, \wtd K_1 &=& \fft{327689\alp\, \ell^2\, r( r^6 - 20r^4\,\ell^2 +
237r^2\, \ell^4 + 118\ell^6)}{243(r^2-\ell^2)^8}\,,\\
h\, \wtd K_2 &=&
\fft{81920\alp\, \ell^2\, (r^6 + 14r^5\,\ell + 35 r^4\, \ell^2 + 
64r^3\,\ell^3
-77 r^2\, \ell^4 + 66 r\, \ell^5 - 39\ell^6)}{243 (r-\ell)^7\,
(r+\ell)^8}\,,\nn
\eea
with $K_3=K_2$ and $\wtd K_3=\wtd K_2$. The $\alp$ correction to the
metric of $G_2$ holonomy could now be worked out explicitly as in the
previous example.

\subsection{$\CP^3$ principal orbits}

 The simplest description of these metrics is the one given in
\cite{cglpnoslump}, in which the left-invariant 1-forms 
$L_{AB}=-L_{BA}$
of $SO(5)$, satisfying $dL_{AB}= L_{AC}\wedge L_{CB}$, are decomposed 
into
\be
R_i= \ft12 (L_{0i} + \ft12 \ep_{ijk}\, L_{jk})\,,\qquad
L_i= \ft12 (L_{0i} - \ft12 \ep_{ijk}\, L_{jk})\,,\qquad
P_a= L_{a4}\,.
\ee
Here the index $0\le A\le 4$ is split as $A=(a,4)=(0,i,4)$, and so 
the $L_i$
and $R_i$ are left-invariant 1-forms of the $SO(4)\sim SU(2)_L\times 
SU(2)_R$ subalgebra, and $P_a$ are in the coset $S^4 = SO(5)/SO(4)$.  
Thus the 1-forms $(R_i,P_a)$ span $S^7$ described as an $SU(2)$ bundle
over $S^4$, and so $(R_1,R_2,P_a)$ span $\CP^3 = S^7/U(1)$, viewed as
an $S^2$ bundle over $S^4$.  The ansatz for cohomogeneity-one metrics
with $\CP^3$ principal orbits is then
\bea
ds^2 = dt^2 + a^2\, P_a^2 + b^2\, (R_1^2 + R_2^2)\,,
\eea
where $a$ and $b$ are functions of $t$.

Following the same procedure as in Section \ref{s3s31sec}, we
find that the first-order equations for $a$ and $b$ are
\be
\fft{\dot a}{a} + \fft{b}{2a^2} +  \dot S_1 = 0\, ,\qquad 
\fft{\dot b}{b} - \fft{b}{2a^2} + \fft{2}{b} +
\,\dot S_2 = 0\,,\label{fo2}
\ee
where $S_1$ and $S_2$ are given by
\bea
S_1 &=& -\ft9{32} 
\alp\, a^{-12}\, (4a^2-b^2) (256 a^4 - 234 a^2\, b^2 +  51 b^4)
\,,\nn\\
S_2 &=& \ft{9}{16} 
\alp\, a^{-12}\, (8a^2-3 b^2) ( 44 a^4 - 38 a^2\, b^2  +9 b^4)
\,.
\label{s1s2sol3}
\eea
These equations can be solved by introducing a new radial variable
$\rho$, defined by $dt=- b^{-1}\, d\rho$.  After analogous
manipulations to those described in Section \ref{s3s31sec}, we find
that $a$ and $b$ are given by
\bea
a^2(\rho) &=& e^{-2S_1(\rho)}\, \int_{\rho_1}^\rho e^{2 S_1(x)}\, 
dx\,,\nn\\
b^2(\rho) &=& \fft{4}{a^2(\rho)}\, e^{-2\left[S_1(\rho) + 
S_2(\rho)\right]}\, 
\int_{\rho_0}^\rho a^2(x)\, e^{2 \left[S_1(x) +  S_2(x)\right]}\, 
dx\,.
\label{cp3sol}
\eea
As in Section \ref{s3s31sec}, we should take $\rho_1<\rho_0$, 
with the simplest choice being such that the expression for $a^2$ in
(\ref{cp3sol}) is given by (\ref{asimpsol}).
Again, these exact integro-differential equations can be solved
easily, at order ${\a'}^3$, since one can substitute the leading-order
solutions $a^2=\rho$, $b^2=2\rho(1-\rho_0^2/\rho^2)$ of the 
$G_2$-holonomy metric (as found in \cite{brysal,gibpagpop}) into the 
expressions (\ref{s1s2sol3}) for $S_1$ and $S_2$.  We find that
\bea 
a^2 &=& \rho + \left(9/16\right) \alp \, \rho^{-8} \left(-24\rho^6 + 
130\rho^4\, \rho_0^2+
616\rho^2\, \rho_0^4 + 459\rho_0^6\right) \nn\\
b^2 &=&2\rho\,\left[1-\left(\rho_0^2/\rho^2\right)\right] + 
\left(9/8\right) \alp 
\left(\rho-\rho_0\right)^2 \rho_0^{-1}\rho^{-10} \Big(326 \rho^7 +
628\rho^6\,\rho_0 + 930\rho^5\,\rho_0^2\nn\\
&& + 1090\rho^4\,\rho_0^3 + 1250\rho^3\,\rho_0^4 +
1108 \rho^2\,\rho_0^5 +966\rho\,\rho_0^6 + 483\rho_0^7\Big)\,.
\eea

\subsection{$SU(3)/U(1)^2$ flag-manifold principal orbits}

   In this example, it is convenient, as in \cite{cglpnoslump} to
introduce the (complex) left-invariant 1-forms $L_A{}^B$ of $SU(3)$,
which satisfy $(L_A{}^B)^\dagger = L_B{}^A$, $L_A{}^A=0$ and 
$dL_A{}^B=
\im\, L_A{}^C\wedge L_C{}^B$, where $A=1,2,3$.  The six real 1-forms
defined by
\be
\sigma_1 + \im\, \sigma_2\equiv L_1{}^3\,,\quad
\Sigma_1 + \im\, \Sigma_2\equiv L_2{}^3\,,\quad
\nu_1 + \im\, \nu_2\equiv L_1{}^2\,,
\ee
span the coset $SU(3)/(U(1)\times U(1))$ of the flag manifold.  

\subsubsection{Biaxial ansatz}

First, we consider a biaxial ansatz for the metric, with
\bea
ds^2 = dt^2 
+ a^2\, (\sigma_1^2 + \sigma_2^2 + \Sigma_1^2 + \Sigma_2^2)
+ b^2 (\nu_1^2 + \nu_2^2) \,,\label{biaxialflag}
\eea
where $a$ and $b$ are functions of $t$.  

Following the same procedures as in the previous examples, we find 
that
the first-order equations for the existence of a singlet Killing 
spinor 
are given by
\be
\fft{\dot a}{a} - \fft{b}{a^2} + \dot S_1 = 0\,, \qquad 
\fft{\dot b}{b} + \fft{b}{a^2} - \fft{2}{b} +  \dot S_2 = 0\,,
\ee
where $S_1$ and $S_2$ are given by
\bea
S_1 &=& -54 \alp\, 
a^{-12}\, (2 a^2 -b^2) ( 20 a^4 - 39 a^2\,b^2 + 17 b^4)\,,\nn\\
S_2 &=& 108 \alp\, a^{-12}\, (20 a^6 -41 a^4\, b^2 + 31 a^2\, b^4 
-9b^6)
\,.
\eea
Following analogous procedures to those applied above, we define a 
new 
radial coordinate here by $dt= b^{-1}\, d\rho$, leading to the 
solution
\bea
a^2(\rho) &=& 2 e^{-2S_1(\rho)}\, \int_{\rho_1}^\rho 
e^{2 S_1(x)}\, dx\,,\nn\\
b^2(\rho) &=& \fft4{a^2(\rho)}\, e^{-2\left[S_1(\rho) + 
S_2(\rho)\right]}\, 
\int_{\rho_0}^\rho a^2(x)\, e^{\left[2 S_1(x) + S_2(x)\right]}\, dx\,.
\eea
As in the previous examples, we can choose $\rho_1$ so that we here 
have
\be
a^2(\rho) = 2 e^{-2S_1(\rho)}\,\left[\rho +  \int_{\rho}^\infty 
\left(1- e^{2 S_1(x)}\right)\, dx \right]. 
\ee
In this case the solution to order $\alp$ is
\bea
a^2 &=& 2 \rho + \left(9/8\right) \alp \left(-72 \rho^6 + 90 
\rho^4\,\rho_0^2
+ 616 \rho^2\, \rho_0^4 - 459 \rho_0^6\right)\, ,\nn\\
b^2 &=& 2\rho\left[1 - \left(\rho_0^2/\rho^2\right)\right] + 
\left(9/8\right) \alp
\left(\rho-\rho_0\right)^2\rho_0^{-1}\rho^{-10} \Big(
502 \rho^7 + 932 \rho^6\, \rho_0 +
1362\rho^5\,\rho_0^2\nn\\
&& \qquad +\,  1306\rho^4\,\rho_0^3 +1250 \rho^3\,\rho_0^4 +
1108 \rho^2\,\rho_0^5 + 966 \rho\,\rho_0^6 + 483 \rho_0^7\Big)\,.
\eea

\subsubsection{Triaxial ansatz}

   The biaxial Flag metric ansatz (\ref{biaxialflag}) can be 
generalised to 
a triaxial ansatz, with
\be
ds^2 = dt^2
+ a^2\, (\sigma_1^2 + \sigma_2^2) + b^2(\Sigma_1^2 + \Sigma_2^2)
+ c^2 (\nu_1^2 + \nu_2^2) \,,
\ee
We find that the first-order equations are
\be
\fft{\dot a}{a} =  \fft{a^2 - b^2 - c^2}{a\,b\,c} - \dot S_1
\ee
and its cyclic permutations with $S_1,S_2,S_3$, where
\bea
S_1 &=& 
-\fft{54 (a^2 + b^2 - c^2)(a^2 - b^2 + c^2)}{a^6\, b^6\, c^6}
\Bigg[ 18a^8 + 17 b^8 + 17 c^8\nn\\
&&-31 a^6\, b^2 - 29 a^2\, b^6 - 31 a^6\,
c^2 - 10 b^6\, c^2 - 29 a^2\, c^6
- 10 b^2 c^6 + 25 a^4\, b^4\nn\\
&& + 25 a^4\, c^4 - 14 b^4\, c^4 +
32 a^4\, b^2\, c^2 + 9 a^2\, b^4\, c^2 + 25 a^2\, b^2\, c^4 
\Bigg]
\eea
and cyclic permutations for $S_2$ and $S_3$.  These equations may be
simplified by defining
\be
u_1 =  b\, c\,,\qquad u_2 = c\, a\,,\qquad u_3= a\, b\,,
\ee
and introducing the new radial coordinate $\rho$ such that $d\rho =
-abc\, dt$.  The equations then become
\be
u_1' -  \fft{2}{u_1} = - u_1\, \left(S_2' + S_3'\right) 
\ee
and cyclic permutations, where the prime indicates differentiation
with respect to $\rho$. These equations can be integrated to give
\be
u_1^2(\rho) = 4 e^{-2S_2(\rho) - 2 S_3(\rho)}\, \int_{\rho_1}^\rho
e^{2 S_2(x) + 2 S_3(x)}\, dx
\ee
and cyclic permutations. The functions $u_!,u_2,u_3$, and hence the
metric coefficients $a,b,c$ may then be found to order $\alp$ as
before.

\section{Lifting to M-theory}

So far we have focused on the $\alp$ corrections at tree level in type
II string theory, which are identical for type IIA and type IIB. The
IIA string can be viewed as an $S^1$ compactification of M-theory, and
to this extent our results are also of direct relevance in
M-theory. However, all tree-level $\alpha'$ corrections to the string
effective action vanish in the decompactification limit in which we
recover uncompactified M-theory.  Nevertheless, there {\it are} $R^4$
corrections in M-theory, which arise in the context of IIA string
theory as {\it one-loop} corrections to the $\alp R^4$ term in the
effective action.  The IIB contribution to the $R^4$ term at one loop
has the same structure as the tree-level term, as required by
$Sl(2;Z)$ duality, but the IIB and IIA contributions are no longer
identical at one-loop because of the different Ramond-Ramond sectors
circulating in the loop. The situation can be summarised as follows
(see, for example, \cite{ferrara,tseytlin}). From (\ref{Y}) one sees
that $Y$ can be written as
\be
Y= Y_0 - E_8
\ee
where $Y_0 \sim t_8t_8 R^4$ and $E_8$ is the 10-dimensional scalar
that reduces to the Euler density on $M_8$. The $\alp$ contributions
to the type IIA and type IIB effective Lagrangians at tree-level and
1-loop can now be summarised by the following table, where CS stands
for a $B\wedge t_8 R^4$ Chern-Simons term \cite{vafwit}:

\begin{table}[ht]
\centering
\caption{String tree-level and one-loop 
corrections\label{tab:treeloopcorrs}}
\vspace{0.4cm}
\begin{tabular}{|c|c|c|c|}\hline
& Tree Level&  One Loop\\ \hline\hline
IIA: & $e^{-2\phi}\, \left(Y_0 - E_8\right)$
& $Y_0 + E_8$ +\ CS
\\ \hline
IIB: & $ e^{-2\phi}\, \left(Y_0- E_8\right)$ &
$Y_0- E_8$ \\ \hline
\end{tabular}
\end{table}

   The one-loop CS contribution to the IIA theory lifts to a similar
$A\wedge t_8 R^4$ term in eleven dimensions
\cite{dulimi}. Fortunately, this term is irrelevant to $G_2$
compactifications, as is the $E_8$ term, in view of their $n=7$ 
dimensionality. 
For our present purposes, we may consistently set the 4-form field 
strength to
zero, in which case the relevant part of the (bosonic) M-theory 
effective
Lagrangian density can therefore be taken to be just
\be\label{lag11} {\cal
L}_{11} = \sqrt{-\hat g}\left[ \hat R - \lambda\, \hat Y_0\right]
\,,\label{d11covcorr}
\ee 
where $\lambda \sim\ell_{11}^6$ must be small in comparison to the 
length scale of the
compactifying solution. It should be understood that the hatted 
quantities are eleven
dimensional.  Of course, this also implies a particular choice of 
field variables; one could also
include terms that vanish by use of the leading-order field equation 
$\hat R_\sst{MN}=0$.
Clearly, the choice of whether to include such terms or not is a 
matter of taste, related to the
desired form of the resulting field equations. Indeed, in the 
following we shall make a specific
choice of field redefinition that has the virtue of producing the 
same form of the internal space
deformations as we obtained previously in Section \ref{g2sec} at 
string tree level. We recall
that, in particular, this has the feature that for an internal space 
of the form $K_7=S^1\times
K_6$, the initially Ricci-flat K\"ahler space $K_6$ remains K\"ahler 
even after the quantum
deformation.\footnote{This scheme choice can be contrasted with that 
made in Ref.\
\cite{froltsey}, which does not preserve the K\"ahler condition.}

In order to apply our previous results directly in eleven dimensions, 
we now consider this field
redefinition in detail.  The need for this redefinition arises from 
the fact that in our string
theory discussion of Section \ref{g2sec}, the eventual form 
(\ref{correctedg2}) for the
modification of the Ricci-flatness condition on the $G_2$ manifold 
included a contribution from an
$\alp$ correction to the dilaton.  In eleven dimensions, where there 
is no dilaton,
the  equations of motion that follow from (\ref{lag11}) would 
therefore
imply a different equation from (\ref{correctedg2}) for the corrected 
$G_2$
metric. We can compensate for the absence of the dilaton by adding a 
term
proportional  to $\hat R\, \hat Z$, where $\hat Z$ is given by 
(\ref{zexp})
evaluated in eleven dimensions, achieved by performing the field 
redefinition
\be
\hat g_\sst{MN}\longrightarrow (1 + \ft29 \lambda\, \hat R\, \hat 
Z)\, 
\hat g_\sst{MN}\,.
\ee
After this field redefinition, the Lagrangian density (\ref{lag11}) 
becomes
\bea
\label{lag11a} {\cal
L}_{11} &=& \sqrt{-\hat g}\left[ \hat R - \lambda\, \hat Y_0 + 
\lambda\, \hat R\, \hat Z\right] \,,\nn\\
&=&  \sqrt{-\hat g}\left[ \hat R - \lambda\, (\hat W_0 - \hat W_1 
-2\hat W_2)
\right]\,.
\eea 
It easily verified that the resulting field equations are solved by
backgrounds of the form (Minkowski)$_4\times K_7$, where $K_7$ has a
modified $G_2$ metric satisfying the same equation (\ref{correctedg2})
as we had for the (Minkowski)$_3 \times K_7$ metrics in string theory.
Note that since $\lambda$ is itself ``small,'' and $\hat R$ is of 
order
$\lambda$, when varying the
additional term in (\ref{lag11a})
we need only retain the terms coming from $\delta \hat R =
(-\hat \nabla_\sst{M}\, \hat \nabla_\sst{N} + \hat g_\sst{MN}\, \hat
{\square})\, \delta \hat g^\sst{MN}+\cdots $. 

The modified $D=11$ equations following from (\ref{lag11a}) with 
leading-order
correction terms specialised to the (Minkowski)$_4 \times K_7$ case 
take the form
\bea
\label{d11correqs}
\hat R_{\mu\nu}-\ft12\hat R\hat g_{\mu\nu} &=& -\lambda 
g_{\mu\nu}\square Z\label{d11corrmunu}\\
\hat R_{ij}-\ft12\hat R\hat g_{ij} &=& 
\lambda(X_{ij}+\nabla_i\nabla_jZ - 
g_{ij}\square Z)\,,\label{d11corrij}
\eea
where $X_{ij}$ was given in Eq.\ (\ref{Xdef}). For $D=11$ spacetimes 
of the form
(Minkowski)$_4 \times K_7$, we have $\hat R=R$, where $R$ is the 
7-dimensional Ricci scalar. 
Using (\ref{d11corrij}) and (\ref{Xtrace}), one has 
$R=2\lambda\square Z$ and hence
\be
R_{ij}=\lambda(X_{ij}+\nabla_i\nabla_j Z)\,. \label{Rcorrij}
\ee
Eq.\ (\ref{d11corrmunu}) is then identically satisfied for a 
(Minkowski)$_4 \times K_7$ spacetime,
because it simply reduces to $-\lambda\square Z\eta_{\mu\nu} = 
-\lambda\square Z\eta_{\mu\nu}$.
Accordingly, the M-theory deformation of a $G_2$ internal manifold as 
given in Eq.\
(\ref{Rcorrij}) is exactly the  same as that given by 
(\ref{correctedg2}) for the $D=10$ string
tree-level case.

   If one uses the standard Kaluza-Klein ansatz for reduction to 
$D=10$
\be
d\hat s^2_{11} = e^{-{2\over3}\tilde\phi} ds^2_{10} + 
e^{{4\over3}\tilde\phi}\, dy^2\,,\label{kkred}
\ee
then one finds the Lagrangian density  
\be
{\cal L}_{10} = \sqrt{-g}\left\{e^{-2\tilde\phi}\left[R +
4\left(\partial\tilde\phi \right)^2\right] - \lambda \,Y_0 
  +\lambda\, Z\, (R + 4\,\square\td\phi)\right\} +\cdots\,,
\label{d10lag1}
\ee
where the ellipses represent terms of higher order in the small 
parameter
$\lambda$, which are unimportant at the order to which we are working.
The trickiest part of this calculation is the evaluation of the
term $\square\td\phi\, Z$, which arises from the Kaluza-Klein 
reduction
of $\hat Y_0$ and $\hat W_2 = \hat R\, \hat Z$.  It is useful to note 
that at 
the linearised level, which suffices for our purposes, the reduction
of the Riemann tensor gives
\be
\hat R_{\mu\nu\rho\sigma} = e^{\ft23\td\phi}\, R_{\mu\nu\rho\sigma} 
  -\ft13 (\nabla_\rho\nabla_\nu\td\phi\, g_{\mu\sigma} -
\nabla_\sigma\nabla_\nu\td\phi\, g_{\mu\rho} -
  \nabla_\rho\nabla_\mu\td\phi\, g_{\nu\sigma} +
  \nabla_\sigma\nabla_\mu\td\phi\, g_{\nu\rho})\,,
\ee
and that the substitution of this into $\hat Y_0$ can be evaluated 
using the variational formula
(\ref{xvar8}) and the relation (\ref{Xtrace}) which holds in the 
original $G_2$
background.

    The terms involving $Z$ in (\ref{d10lag1}) can then be removed by 
redefining the dilaton according to
\be
\td\phi = \phi +\ft12 \lambda\, Z\,,\label{phiredef}
\ee
which implies that (\ref{d10lag1}) becomes
\be
{\cal L}_{10} = \sqrt{-g}\left\{e^{-2\phi}\left[R +
4\left(\partial\phi \right)^2\right] - \lambda \, Y_0 \right\} 
+\cdots\,,
\label{d10lag2}
\ee
This is the part of the $\alp$ corrected effective action given in 
Table
\ref{tab:treeloopcorrs} at one string loop for the type IIA theory 
that is relevant
to $G_2$ compactifications.\footnote{The tree-level
contribution arises from consideration of the Kaluza-Klein modes
\cite{green}.}  The dilaton redefinition (\ref{phiredef}) is 
easily understandable when one recalls that the field $\td\phi$ 
appearing
in the Kaluza-Klein reduction ansatz (\ref{kkred}) is constant in the
(Minkowski)$_4\times K_7$ background, whereas the redefined dilaton
$\phi$ appearing in (\ref{phiredef}) is non-constant, and precisely
reproduces what we found in (\ref{correctdilaton}) in the 
string compactification discussion of Section \ref{g2sec}. 

If one tried to oxidise directly the $D=10$ corrected Lagrangian 
${\cal L}_{10}$ of Eq.\
(\ref{d10lag2}) back to $D=11$ via the standard Kaluza-Klein ansatz 
$d\hat s^2_{11} =
e^{-{2\over3}\phi} ds^2_{10} +  e^{{4\over3}\phi}\, dy^2$, one would
not obtain a manifestly $D=11$ covariant result, since  $g_{11\,11}$ 
would appear asymmetrically in
the correction terms. Moreover, the (Minkowski)$_3\times K_7$ vacuum 
of the $D=10$ theory would
not lift via this oxidation up to a manifestly $D=4$ Lorentz 
invariant (Minkowski)$_4\times K_7$
solution. The field redefinition (\ref{phiredef}) resolves these 
issues, allowing one to pass
between the quantum corrected $D=11$ covariant action (\ref{lag11a}) 
and the corrected $D=10$
string action (\ref{d10lag2}), and similarly for the 
(Minkowski)$\times K_7$ solutions. Of course,
one could equivalently wrap the field redefinition (\ref{kkred}) into 
the Kaluza-Klein ansatz and
so view the passage between the
$D=11$ and $D=10$ theories to be via a quantum corrected KK ansatz.

We have thus shown (in close analogy to string theory) that the $R^4$ 
corrections to M-theory
preserve the supersymmetry of the Kaluza-Klein vacua of 
11-dimensional supergravity of the
form (Minkowski)$_4\times K_7$ in which $K_7$ is a (Ricci flat) 
7-manifold of G2 holonomy,
although (as in $D=10$ string theory) these corrections do not in 
fact preserve either the Ricci
flatness of $K_7$ or its $G_2$ holonomy. This circumstance may 
suggest that an appropriate
perspective for interpreting the preservation of unbroken 
supersymmetry might be that of generalised
holonomies \cite{genhol}.

\section*{Acknowledgments}

   We would like to thank Michael Green, Dominic Joyce, Kasper 
Peeters, Arkady
Tseytlin, Pierre Vanhove and Anders Westerberg for helpful 
discussions and 
correspondence.
C.N.P. and K.S.S. thank the University of Barcelona, and C.N.P. 
thanks 
the General Relativity group at the University of Cambridge, for
hospitality during the course of this work.

\end{document}